\documentclass[12pt,thmsa]{article}
\usepackage{sw20lart}

\input{tcilatex}
\begin{document}

\author{Emilio Santos \and Departamento de F\'{i}sica. Universidad de Cantabria.
Santander. Spain}
\title{The relevance of random choice in tests of Bell inequalities with two atomic
qubits}
\maketitle

\begin{abstract}
It is pointed out that a loophole exists in experimental tests of Bell
inequality using atomic qubits, due to possible errors in the rotation
angles of the atomic states. A sufficient condition is derived for closing
the loophole.

PACS numbers: 03.65.Ud, 03.67.Mn, 37.10.Ty, 42.50.Xa
\end{abstract}

A recent experiment, by a group of Maryland, has measured the correlation
between the quantum states of two Yb$^{+}$ ions separated by a distance of
about 1 meter\cite{Mats}. The authors claim that the experiment is relevant
because it violates a CHSH\cite{CHSH} (Bell) inequality, modulo the locality
loophole, closing the detection loophole. In my opinion that assertion does
not make full justice to the relevance of the experiment. The truth is that
it is the first experiment which has tested a genuine Bell inequality.
Actually the results of previous experiments, in particular those involving
optical photon pairs\cite{Genovese}, did not test any \textit{genuine }Bell
inequality, that is an inequality which is a necessary condition for the
existence of local hidden variables (LHV) models. The inequalities tested in
those experiments should not be qualified as Bell\'{}s because their
derivation involves additional assumptions. Consequently their violation
refutes only restricted families of LHV models, namely those fulfilling the
additional assumption. ( For details see\cite{Santos}.)

The aim of the present letter is to point out the existence of a loophole in
the Maryland experiment\cite{Mats}, or more generally in Bell tests with
atomic qubits, in addition to the locality loophole. Blocking that loophole
will be straightforward using random choice of the measurements, as is
explained below.

In general I will consider experiments where a pair of atoms (or ions) is
prepared in an entangled state. Then Alice performs a rotation of the state
of her atom by an angle $\theta _{a}$ and, after a short time, she may
detect fluorescence of the atom illuminated by an appropriate laser.
Similarly Bob performs a rotation of his atom by an angle $\theta _{b}$ and,
after that, he may detect fluorescence too. I shall label $p_{++}\left(
\theta _{a},\theta _{b}\right) $ the probability of coincidence detection
and $p_{--}\left( \theta _{a},\theta _{b}\right) $ the probability that
neither Alice nor Bob detect fluorescence. Similarly $p_{-+}\left( \theta
_{a},\theta _{b}\right) $ ( $p_{+-}\left( \theta _{a},\theta _{b}\right) )$
will be the probability that only Bob (Alice) detects fluorescence. In the
Maryland experiment\cite{Mats} (see their eq.$\left( 6\right) )$, a function 
$E\left( \theta _{a},\theta _{b}\right) $ is defined by 
\begin{equation}
E\left( \theta _{a},\theta _{b}\right) =p_{++}\left( \theta _{a},\theta
_{b}\right) +p_{--}\left( \theta _{a},\theta _{b}\right) -p_{+-}\left(
\theta _{a},\theta _{b}\right) -p_{-+}\left( \theta _{a},\theta _{b}\right) .
\label{a7}
\end{equation}
Then the authors define a parameter $S$ by 
\begin{equation}
S=\left| E\left( \theta _{a},\theta _{b}\right) +E\left( \theta _{a}^{\prime
},\theta _{b}\right) \right| +\left| E\left( \theta _{a},\theta _{b}^{\prime
}\right) -E\left( \theta _{a}^{\prime },\theta _{b}^{\prime }\right) \right|
,  \label{7a}
\end{equation}
and claim that the CHSH\cite{CHSH} inequality $S\leq 2$ is violated. The
notation used by the authors is, however, somewhat misleading. Instead of eq.%
$\left( \ref{a7}\right) $ they write

\begin{equation}
E\left( \theta _{a},\theta _{b}\right) =p\left( \theta _{a},\theta
_{b}\right) +p\left( \theta _{a}+\pi ,\theta _{b}+\pi \right) -p\left(
\theta _{a},\theta _{b}+\pi \right) -p\left( \theta _{a}+\pi ,\theta
_{b}\right) ,  \label{10}
\end{equation}
where they label $p\left( \theta _{a},\theta _{b}\right) $ the quantity
which I have labeled $p_{++}\left( \theta _{a},\theta _{b}\right) .$
Definition eq.$\left( \ref{10}\right) ,$ in place of eq.$\left( \ref{a7}%
\right) ,$ rests upon assuming the equalities 
\begin{eqnarray*}
p_{-+}\left( \theta _{a},\theta _{b}\right) &=&p\left( \theta _{a}+\pi
,\theta _{b}\right) ,p_{+-}\left( \theta _{a},\theta _{b}\right) =p\left(
\theta _{a}+\pi ,\theta _{b}\right) , \\
p_{--}\left( \theta _{a},\theta _{b}\right) &=&p\left( \theta _{a}+\pi
,\theta _{b}+\pi \right) ,
\end{eqnarray*}
which are true according to quantum mechanics, but may not be true in LHV
theories. In any case the authors measured $E\left( \theta _{a},\theta
_{b}\right) $ as defined in eq.$\left( \ref{a7}\right) $\cite{priv}.

In order to show that there is a loophole in the experiment, in addition to
the locality loophole, I begin remembering that, according to Bell\cite{Bell}%
, a LHV model will contain a set of hidden variables, $\lambda ,$ a positive
normalized density function, $\rho \left( \lambda \right) ,$ and two
functions $M_{a}\left( \lambda ,\theta _{a}\right) ,$ $M_{b}\left( \lambda
,\theta _{b}\right) $, $\theta _{a}$ and $\theta _{b}$ being parameters
which may be controlled by Alice and Bob respectively. The latter functions
fulfil 
\begin{equation}
M_{a}\left( \lambda ,\theta _{a}\right) ,M_{b}\left( \lambda ,\theta
_{b}\right) \in \{0,1\}.  \label{1b}
\end{equation}
In the Maryland experiment\cite{Mats} the parameters $\theta _{a}$ and $%
\theta _{b}$ are angles defining the quantum states of the two ions. The
probability, $p_{++}\left( \theta _{a},\theta _{b}\right) ,$ that the
coincidence measurement of two dichotomic variables, in two distant regions,
gives a positive answer for both variables should be obtained in the LHV\
model by means of the integral 
\begin{equation}
p_{++}\left( \theta _{a},\theta _{b}\right) =\int \rho \left( \lambda
\right) M_{a}\left( \lambda ,\theta _{a}\right) M_{b}\left( \lambda ,\theta
_{b}\right) d\lambda .  \label{1}
\end{equation}
Similarly the probability, $p_{+-}\left( \theta _{a},\theta _{b}\right) ,$
that Alice gets the answer ``yes'' and Bob the answer ``no'' is given by 
\begin{equation}
p_{+-}\left( \theta _{a},\theta _{b}\right) =\int \rho \left( \lambda
\right) M_{a}\left( \lambda ,\theta _{a}\right) \left[ 1-M_{b}\left( \lambda
,\theta _{b}\right) \right] d\lambda ,  \label{1a}
\end{equation}
and analogous expressions for $p_{-+}$ and $p_{--}.$

A LHV model for an atomic experiment may be obtained by choosing 
\begin{eqnarray}
\rho \left( \lambda \right) &=&\frac{1}{2\pi },\lambda \in \left[ 0,2\pi
\right] ,\;M_{a}\left( \lambda ,\theta _{a}\right) =\Theta \left( \frac{\pi 
}{2}-\left| \lambda -\theta _{a}\right| \right) ,  \nonumber \\
M_{b}\left( \lambda ,\theta _{b}\right) &=&\Theta \left( \frac{\pi }{2}%
-\left| \lambda -\theta _{b}-\pi \right| \right) ,\func{mod}\left( 2\pi
\right) ,  \label{31}
\end{eqnarray}
where $\Theta \left( x\right) =1$ if $x>0$, $\Theta \left( x\right) =0$ if $%
x<0$. It is easy to see, taking eqs.$\left( \ref{1}\right) $ and $\left( \ref
{1a}\right) $ into account, that model predictions are (assuming $\theta
_{a},\theta _{b}\in \left[ 0,\pi \right] )$%
\begin{eqnarray}
p_{++}\left( \theta _{a},\theta _{b}\right) &=&p_{--}\left( \theta
_{a},\theta _{b}\right) =\frac{\left| \theta _{a}-\theta _{b}\right| }{2\pi }%
,  \nonumber \\
p_{+-}\left( \theta _{a},\theta _{b}\right) &=&p_{-+}\left( \theta
_{a},\theta _{b}\right) =\frac{1}{2}-\frac{\left| \theta _{a}-\theta
_{b}\right| }{2\pi }.  \label{32}
\end{eqnarray}
Hence I get 
\begin{equation}
E\left( \theta _{a},\theta _{b}\right) =\frac{2}{\pi }\left| \theta
_{a}-\theta _{b}\right| -1,  \label{33}
\end{equation}
and it is not difficult to show that, for any choice of the angles $\theta
_{a},\theta _{b},\theta _{a}^{\prime },\theta _{b}^{\prime },$ the model
predicts $S\leq 2$ with $S$ given by eq.$\left( \ref{7a}\right) .$

Now let us assume that the experiment is performed so that Alice and Bob
start measuring the quantity $E\left( \theta _{a},\theta _{b}\right) $ in a
sequence of runs of the experiment. After that they measure $E\left( \theta
_{a},\theta _{b}^{\prime }\right) $ in another sequence, then they measure $%
E\left( \theta _{a}^{\prime },\theta _{b}\right) $ and, finally, they
measure $E\left( \theta _{a}^{\prime },\theta _{b}^{\prime }\right) .$ Let $%
\alpha $ be the error in the rotation performed by Bob on his atom in the
first sequence of runs, so that the rotation angle is $\theta _{b}+\alpha $
rather than $\theta _{b}$ in the measurement of $E\left( \theta _{a},\theta
_{b}\right) .$ Similarly I shall assume that the rotation angles are $\theta
_{b}^{\prime }+\beta ,\theta _{b}+\gamma $ and $\theta _{b}^{\prime }+\delta 
$ in the measurements of $E\left( \theta _{a},\theta _{b}^{\prime }\right) $%
, $E\left( \theta _{a}^{\prime },\theta _{b}\right) $ and $E\left( \theta
_{a}^{\prime },\theta _{b}^{\prime }\right) ,$ respectively. For simplicity
I will assume that no error appears in Alice rotations. The errors are
considered small, specifically $\left| \alpha \right| ,\left| \beta \right|
,\left| \gamma \right| ,\left| \delta \right| <\pi /4.$ I shall prove that,
taking into account the errors in the measurement of the angles, the LHV
model prediction for the parameter $S$, eq.$\left( \ref{7a}\right) $ may
apparently violate the CHSH\cite{CHSH} inequality $S\leq 2.$ To do that let
us choose, as in the Maryland experiment\cite{Mats}, 
\begin{equation}
\theta _{a}=\frac{\pi }{2},\theta _{b}=\frac{\pi }{4},\theta _{a}^{\prime
}=0,\theta _{b}^{\prime }=\frac{3\pi }{4}.  \label{34}
\end{equation}
The values predicted by the LHV model for the relevant quantities are 
\begin{eqnarray}
E\left( \theta _{a},\theta _{b}+\alpha \right) &=&-0.5-\frac{2\alpha }{\pi }%
,\;E\left( \theta _{a},\theta _{b}^{\prime }+\beta \right) =-0.5+\frac{%
2\beta }{\pi },  \nonumber \\
E\left( \theta _{a}^{\prime },\theta _{b}+\gamma \right) &=&-0.5+\frac{%
2\gamma }{\pi },\;E\left( \theta _{a}^{\prime },\theta _{b}^{\prime }+\delta
\right) =0.5+\frac{2\delta }{\pi }.  \label{35}
\end{eqnarray}
Then the parameter actually measured in the experiment is 
\begin{equation}
S^{\prime }=\left| E\left( \theta _{a},\theta _{b}+\alpha \right) +E\left(
\theta _{a}^{\prime },\theta _{b}+\gamma \right) \right| +\left| E\left(
\theta _{a},\theta _{b}^{\prime }+\beta \right) -E\left( \theta _{a}^{\prime
},\theta _{b}^{\prime }+\delta \right) \right| ,  \label{46}
\end{equation}
and the LHV prediction for that parameter is 
\[
S^{\prime }==2+\frac{2}{\pi }\left( \alpha -\beta -\gamma +\delta \right) , 
\]
which may violate the inequality $S^{\prime }\leq 2$ for some values of the
parameters $\alpha ,\beta ,\gamma $ and $\delta .$ In particular the results
of the Maryland experiment\cite{Mats} are reproduced by choosing 
\[
2\alpha /\pi =0.018,2\beta /\pi =-0.046,2\gamma /\pi =-0.081,2\delta /\pi
=-0.073. 
\]
The errors in the angles are of order 7${{}^{o}}$ or less. It is plausible
that errors as high as these may appear in experiments with atomic qubits
but not in optical photon experiments. I stress that no violation of a Bell
inequality by a\ LHV model is produced. Actually the parameter $S^{\prime }$
of eq.$\left( \ref{46}\right) $ is not a CHSH parameter as defined in eq.$%
\left( \ref{7a}\right) .$

In the following I shall prove that the loophole may be closed by random
choice of the angles to be measured. To begin with, it is easy to see that
the LHV model predictions do not violate the inequality $S^{\prime }\leq 2$
if the error in the measurement, by Bob, of the angle $\theta _{b}$ is the
same in all measurements of that angle, and similarly for $\theta
_{b}^{\prime }.$ In fact the inequality is fulfilled if $\alpha =\beta $ and 
$\gamma =\delta ,$ as may be seen by looking at eq.$\left( \ref{46}\right) .$
In the following I derive a sufficient condition for the fulfillement of the
inequality, $S^{\prime }\leq 2,$ for the actually measurable quantity $%
S^{\prime },$ by the predictions of any LHV model.

Let us assume that there is a (normalized) probability distribution, $%
f_{a}(x),$ for the errors when Alice rotates her atom by an angle $\theta
_{a}$ and another distribution, $f_{a}^{\prime }(y),$ when she rotates her
atom by an angle $\theta _{a}^{\prime }.$ Similarly I shall assume that
there are similar disitribuions $f_{b}(u)$ and $f_{b}^{\prime }(v)$ for the
errors in the rotations, by Bob, of the angles $\theta _{b}$ and $\theta
_{b}^{\prime }.$ I shall show that a sufficient condition for the inequality 
$S^{\prime }\leq 2$ is that the distributions of errors, in the rotations
made by Alice, are the same independently of what rotation performs Bob on
the partner atom. And similarly for the rotations made by Bob. If this is
the case the predictions of any LHV model for the quantity $S^{\prime }$
will be obtained from probabilities defined as follows (compare with eqs.$%
\left( \ref{1}\right) $ and $\left( \ref{1a}\right) )$%
\begin{eqnarray}
p_{++}\left( \theta _{a},\theta _{b}\right) &=&\int \rho \left( \lambda
\right) M_{a}\left( \lambda ,\theta _{a}+x\right) M_{b}\left( \lambda
,\theta _{b}+u\right) d\lambda f_{a}(x)dxf_{b}(u)du,  \label{37} \\
p_{+-}\left( \theta _{a},\theta _{b}\right) &=&\int \rho \left( \lambda
\right) M_{a}\left( \lambda ,\theta _{a}+x\right) \left[ 1-M_{b}\left(
\lambda ,\theta _{b}+u\right) \right] d\lambda f_{a}(x)dxf_{b}(u)du, 
\nonumber
\end{eqnarray}
and similarly for the other quantities $p_{ij}$ with $i,j=+,-$. Now we may
define new quantities 
\begin{eqnarray}
Q_{a}\left( \lambda ,a\right) &=&\int M_{a}\left( \lambda ,\theta
_{a}+x\right) f_{a}(x)dx,  \label{39} \\
Q_{b}\left( \lambda ,b\right) &=&\int M_{b}\left( \lambda ,\theta
_{b}+u\right) f_{b}(u)du,  \nonumber \\
Q_{a}\left( \lambda ,a^{\prime }\right) &=&\int M_{a}\left( \lambda ,\theta
_{a}^{\prime }+y\right) f_{a}^{\prime }(y)dy,  \nonumber \\
Q_{b}\left( \lambda ,b^{\prime }\right) &=&\int M_{b}\left( \lambda ,\theta
_{b}^{\prime }+v\right) f_{b}^{\prime }(v)dv,  \nonumber
\end{eqnarray}
which fulfil the conditions (compare with eqs.$\left( \ref{1b}\right) )$%
\begin{equation}
0\leq Q_{a}\left( \lambda ,a\right) ,Q_{a}\left( \lambda ,a^{\prime }\right)
,Q_{b}\left( \lambda ,b\right) ,Q_{b}\left( \lambda ,b^{\prime }\right) \leq
1.  \label{40}
\end{equation}
The consequence is that we may obtain a new LHV model for the experiment
with the quantities $Q,$ eqs.$\left( \ref{40}\right) ,$ in place of the
quantities $M$, eqs.$\left( \ref{1b}\right) .$ The existence of the model
implies the fulfillement of the inequality $S^{\prime }\leq 2.$

From our proof it is rather obvious that the essential condition required to
block the loophole is that the probability distribution of errors made by
Bob are independent of what rotation is performed by Alice in the partner
atom, and similarly the errors made by Alice should be independent of the
rotation performed by Bob. A simple method to insure that independence is
that Alice makes at random the choice whether to rotate her atom by the
angle $\theta _{a}$ or by the angle $\theta _{a}^{\prime },$ and similarly
Bob. That is, after every preparation of the entangled state of the atom,
Alice should make a random choice (with equal probabilities) between the
rotation angles $\theta _{a}$ and $\theta _{a}^{\prime }$ and similarly Bob
should make a random choice, \textit{independently of Alice}, between $%
\theta _{b}$ and $\theta _{b}^{\prime }.$


\begin{thebibliography}{9}
\bibitem{Mats}  D. N. Matsukevich, P. Maunz, D.L. Moehring, S. Olmschenk and
C. Monroe, \textit{Phys. Rev. Lett.} \textbf{100}, 150404 (2008).

\bibitem{CHSH}  J. F. Clauser, M. A. Horne, A. Shimony and R. A. Holt, 
\textit{Phys. Rev. Lett.} \textbf{23}, 880 (1969).

\bibitem{Genovese}  M. Genovese, \textit{Phys. Reports} \textbf{413}, 319
(2005).

\bibitem{Santos}  E. Santos, Found. Phys. \textbf{34}, 1643 (2004).

\bibitem{priv}  D. N. Matsukevich, private communication.

\bibitem{Bell}  J. S. Bell, Physics \textbf{1}, 195 (1964).
\end{thebibliography}
\end{document}